\documentclass[pre,showpacs]{revtex4}

\usepackage{amsmath}
\usepackage{amsfonts}
\usepackage{amssymb}
\usepackage[dvips]{graphicx}

\newtheorem{theorem}{Theorem}

\newcommand{\tr}{\mathop{\text{tr}}}

\begin{document}

\title{Phantom cosmology as a simple model with dynamical complexity} 
\author{Marek Szyd{\l}owski}
\email{uoszydlo@cyf-kr.edu.pl}
\affiliation{Complex Systems Research Center, Jagiellonian University, 
Reymonta 4, 30-059 Krak{\'o}w, Poland}
\author{Adam Krawiec}
\email{uukrawie@cyf-kr.edu.pl}
\affiliation{Institute of Public Affairs, Jagiellonian University, 
Rynek G{\l}{\'o}wny 8, 31-042 Krak{\'o}w, Poland}
\author{Wojciech Czaja}
\email{czaja@oa.uj.edu.pl}
\affiliation{Astronomical Observatory, Jagiellonian University, 
Orla 171, 30-244 Krak{\'o}w, Poland}

\begin{abstract}
We study the Friedmann-Robertson-Walker model with phantom fields modelled in 
terms of scalar fields. We apply the Ziglin theory of integrability and find 
that the flat model is non-integrable. Then we cannot expect to determine 
simple analytical solutions of the Einstein equations. We demonstrate that 
there is only a discrete set of parameters where this model is integrable. 
For comparison we describe the phantoms fields in terms of the barotropic 
equation of state. It is shown that in the contrast to the phantoms modelled 
as scalar fields, the dynamics is always integrable and phase portraits are 
contracted. In this case we find the duality relation. 
\end{abstract}


\maketitle

\section{Introduction}

The recently available measurements of luminosity distances of the type Ia 
supernova (SNIa) as a function of redshift have shown that current Universe 
is in an accelerating phase due to unknown form of repulsive energy 
\cite{Perlmutter:1998,Riess:1998}. The most popular candidate for this dark 
energy is the cosmological constant. On the other hand the results of 
large-scale structure surveys and results of measurements of masses of 
galaxies give best fit for density parameter for matter 
$\Omega_{\text{m},0}=0.3$ \cite{Szalay:2003,Viana:2003} (for review of 
cosmological parameters see Ref.~\cite{Lahav:2004}). Combining data from SNIa 
with measurements of the cosmic microwave background (CMB) radiation, we obtain 
$\Omega_{\Lambda,0}=0.7$ as the best fit value. The sum of the densities 
$\Omega_{\text{total},0}= 1.02 \pm 0.02$ obtained by the Wilkinson microwave 
anisotropy probe (WMAP) \cite{Spergel:2003} agrees with value predicted by 
inflation and suggests that our Universe is almost flat on the large scales. 
Therefore, the assumption of flat model with the cosmological constant is in 
good agreement with observations. 

The acceleration of the Universe can be explained in two-fold manner. 
In first approach it is postulated that there is some unknown exotic matter 
which violates the strong energy condition $\rho + 3p \geq 0$, where $p$ is 
the pressure and $\rho$ is the energy density of perfect fluid. This form of 
matter is called dark energy. In the past few years different scalar field 
models like quintessence and more recently the tachyonic scalar field have 
been conjectured for modeling the dark energy in terms of sub-negative 
pressure $p > - \rho$. A scalar field with super-negative pressure 
$p < - \rho$ called a phantom field can formally be obtained by switching 
the sign of the kinetic energy in the Lagrangian for a standard scalar field. 
For example in the Friedmann-Robertson-Walker (FRW) model the phantom field 
minimally coupled to a gravity field leads to $\rho + p = - {\dot{\phi}}^{2}$, 
where $\rho_{\phi} = -1/2 \dot{\phi}^{2} + V(\phi)$, 
$p_{\phi} = -1/2 \dot{\phi}^{2} - V(\phi)$, and $V(\phi)$ is the phantom 
potential. Such a field was called the phantom field by Caldwell 
\cite{Caldwell:2002} who proposed it as a possible explanation of the observed 
acceleration of the current universe when $\Omega_{\text{m},0} \gtrsim 0.2$. 
Note that a coupling to gravity in the quintessence models was also explored 
\cite{Uzan:1999}.

The second approach called the Cardassian expansion scenario has recently been 
proposed by Freese and Lewis \cite{Freese:2002} as an alternative to dark 
energy in order to explain the current accelerated expansion of the universe. 
In this scenario universe is flat and matter dominated but the standard FRW 
dynamics is modified by the presence of an additional term $\rho^{n}$ such that 
$3H^{2} = \rho_{\text{eff}} = \rho + 3B \rho^{n}$, where $H = (d\ln{a})/dt$ is 
the Hubble parameter; and $a$ is the scale factor. However, let us note that 
this additional term can be interpreted as a phantom field modelled by the 
equation of state $p = p(\rho) = [n(1+ \gamma) - 1] \rho$, where 
$\rho = \rho_{\text{m},0} a^{-3(1+ \gamma)}$. Therefore for dust matter we 
obtain $p = (n-1) \rho$, and $n<0$ leads to the phantom field.

The phantom scalar field can also be motivated from S-brane in string theory 
\cite{Townsend:2003,Ohta:2003,Roy:2003}. The non-canonical kinetic energy 
also occurs in higher-order theories of gravity and super-gravity 
\cite{Nilles:1984,Pollock:1988}. At first, phantom fields were introduced by 
bulk viscosity effects which can be present in FRW cosmology. They are 
equivalent to effective pressure $p_{\text{eff}}=p-3 \xi H$, where $\xi$ is 
a bulk viscosity coefficient. It is because dissipation in general relativity 
is connected (in contrast to friction in classical mechanics) with creation of 
the energy in the expanding universe by the negative pressure contribution 
\cite{Belinskii77,Szydlowski84}.

Without making some specific assumptions on $w(z)$ it very difficult to 
constrain it from the SNIa data \cite{Maor:2002}. Because the astronomical 
observations do not seem to exclude the phantom fields which violate the weak 
energy condition, it is interesting to investigate the theoretical possibility 
to describe dark energy in terms of a phantom field \cite{Parker:1999,
Boisseau:2000,Schulz:2001,Faraoni:2002,Onemli:2002,Hannestad:2002,
Melchiorri:2003,Carroll:2003,Hao03a,Caldwell:2003,Lima:2003,Nojiri:2003b,
Singh:2003,Dabrowski:2003,Hao:2004b,Johri:2004,Piao:2003,Alam:2004}. 
The Cardassian expansion with $n<0$ which can be interpreted as the phantom 
fluid effect is also statistically admissible from SNIa data observations 
\cite{Choudhury:2005}.

In this paper we ask what kind of dynamics can be expected from the FRW model 
with phantom field. It is well known that the standard FRW model reveals some 
complex dynamics. The detailed studies gave us a deeper understanding of 
dynamical complexity and chaos in cosmological models and resulted in 
conclusion that complex behavior depends on the choice of a time 
parameterization or a lapse function in general relativity \cite{Cornish:1996}. 
Castagnino et al. \cite{Castagnino:2000} showed that dynamics of closed FRW 
models with conformally coupled massive scalar field is not chaotic if 
considered in the cosmological time. They showed that for all initial 
conditions the universe will collapse in finite time and then conclude that 
there is no chaos in the model. In their work the monotonously growing 
function is defined along a trajectory which diverge at infinity for arbitrary 
initial conditions. The same model was analyzed in the conformal time by 
Calzetta and Hasi \cite{Calzetta:1993} who presented the existence of chaotic 
behavior of trajectories in the phase space. 

For the cosmological FRW model with a scalar field the kinetic energy form is 
indefinite, therefore, the domain admissible for motion is $R^n$. The similar 
situation happens in the Bianchi IX in which, as it was proved by Cushman and 
Sniatycki \cite{Cushman:1995}, trajectories have no recurrence property. 

The standard methods of chaos investigation can be also applied to the 
wide class of a relativistic system \cite{Motter:2003jm}. Motter and 
Letelier \cite{Motter:2002} explained that this contradiction in the results 
is obtained because the system under consideration is non-integrable. 
Therefore we can speak about complex dynamics in terms of non-integrability 
rather than deterministic chaos. The significant feature is that 
non-integrability is an invariant evidence of dynamical complexity in general 
relativity and cosmology 
\cite{Maciejewski98,Maciejewski00,Maciejewski01,Maciejewski02}. 

There are different motivations to study noningrability in general 
relativity and cosmology. One reason is the possible physical implications 
of existence of complexity in the systems which for example could help 
to explain the formation of structures. Another reason is to develop 
suitable tools to study relativistic system. The next motivation is to 
understand the ultimate implication of the time reparameterization. 

For the FRW model with phantom it can be shown that there is a monotonous 
function along its trajectories and it is not possible to obtain the Lyapunov 
exponents or construct the Poincar{\'e} sections. Therefore we turn to study 
of non-integrability of the phantom system and set it in a much stronger form 
by proving that the system does not possesses any additional and independent 
of Hamiltonian first integrals, which are in the form of analytic or 
meromorphic functions. Of course, it is not the evidence of sensitive 
dependence of solution on a small change of initial conditions. However, it 
is the possible evidence of complexity of dynamical behavior formulated in 
an invariant way. We study non-integrability in the FRW model with phantom 
fields and find that non-integrability is a generic feature of this model and 
favors rather non-analytical forms of the equation of state. 

It is useful to distinguish between solvability and integrability. 
While integrability is intrinsic property of the system which impose 
the constraints on the solutions in the phase space, the solvability 
is related to the existence of closed form solutions \cite{Goriely:2001}. 
In this paper we concentrate on first integrals rather than 
solutions of a system. 

We study nonintegrability instead of chaos because this criterion is 
invariant with respect to time reparameterization. Note that while this 
program of nonintegrability investigation was explicitly formulated by 
Motter and Letelier \cite{Motter:2002}, this idea was materialized in 
papers by Maciejewski and Szydlowski 
\cite{Maciejewski98,Maciejewski00,Maciejewski01,Maciejewski02}. 
Maciejewski and Szydlowski also showed that the Bianchi VIII and IX are 
noningrable in the sense of non-existence of additional analytic first 
integrals \cite{Maciejewski:1999} and that the Bianchi VIII model is 
non-integrable in the sense of nonexistence of mereomorphic first integrals 
\cite{Maciejewski:2001}. The mereomorphic function possesses only poles as 
its singularities; roughly speaking it is the quotient of analytic functions. 
The latter method is used in this paper. Ziglin proved independently 
non-integrability of FRW closed model with scalar field in the sense of 
nonexistence of additional mereomorphic first integrals \cite{Ziglin:2000}. 
In turn Morales Ruiz and Ramis proved nonintegrability of the Bianchi IX in 
the same sense \cite{Morales:2001b}.

For comparison we consider the FRW model with phantom given by the barotropic 
equation of state which violates the weak energy condition. We obtain that 
this model is integrable in contrast to the previous treatment of phantom 
cosmology. Assuming the barotropic form of the equation of state for the 
phantom model we obtain the integrable dynamics at very beginning.

\section{Hamiltonian dynamics of phantom cosmology}

We assume the model with FRW geometry, i.e., the line element has the form
\begin{equation}
ds^{2}=a^{2}(\eta)[-d\eta^{2}+d\chi^{2}+f^{2}(\chi)(d\theta^{2}
+\sin^{2}{\theta}d\varphi^{2})],
\label{eq:1}
\end{equation}
where
\begin{equation}
f(\chi) = \left \{
\begin{array}{lll} 
\sin{\chi},  & 0 \leq \chi \leq \pi     & k=+1 \\
\chi,        & 0 \leq \chi \leq \infty  & k=0  \\
\sinh{\chi}, & 0 \leq \chi \leq \infty  & k=-1 
\end{array} 
\right.
\label{eq:2}
\end{equation}
$k=0,\pm 1$ is the curvature index, $0 \leq \varphi \leq 2\pi$ and 
$0 \leq \theta \leq \pi$ are comoving coordinates, $\eta$ stands for the 
conformal time such that $dt/a \equiv d\eta$.

It is also assumed that a source of gravity is the phantom scalar field 
$\psi$ with a generic coupling to gravity. The gravitational dynamics is 
described by the standard Einstein-Hilbert action
\begin{equation}
S_{g}=-\frac{1}{2}m_{p}^{2}\int d^{4}x \sqrt{-g}(R-2\Lambda),
\label{eq:3}
\end{equation}
where $m_{p}^{2}=(8\pi G)^{-1}$; for simplicity and without loss of generality 
we assume $4\pi G/3=1$. The action for the matter source is
\begin{equation}
S_{ph}=-\frac{1}{2}\int d^{4}x \sqrt{-g}(-g^{\mu\nu}\psi_{\mu}\psi_{\nu}
+2U(\psi)+\xi R\psi^{2}).
\label{eq:4}
\end{equation}
Let us note that the formal sign of $||\psi||^{2}$ is opposite to that which 
describes the standard scalar field as a source of gravity, where $U(\psi)$ 
is a scalar field potential. We assume
\begin{equation}
U(\psi)=\frac{1}{2}m^{2}\psi^{2}+\frac{1}{4}\lambda\psi^{4}
\label{eq:5}
\end{equation}
and that conformal volume $\int d^{3}x$ over the spatial 3-hypersurface is 
a unit. $\xi$ is a coupling constant of scalar field to the Ricci scalar
\begin{equation}
R=6\left(\frac{\ddot{a}}{a^{3}}+\frac{k}{a^{2}}\right).
\label{eq:6}
\end{equation}

If we have the minimally coupled scalar field then $\xi=0$. We assume a 
non-minimal coupling of the scalar field $\xi \neq 0$. 

The dynamical equation for phantom cosmology in which the phantom field is 
modelled by the scalar field with an opposite sign of the kinetic term in 
action can be obtained from the variational principle 
$\delta(S_{g}+S_{\text{ph}})=0$. After dropping the full derivatives with 
respect to the conformal time we obtain the dynamical equation for phantom 
cosmology from variation $\delta(S_{g}+S_{\text{ph}})/\delta g=0$ as well 
as the dynamical equation for field from variation 
$\delta(S_{g}+S_{\text{ph}})/\delta \psi=0$ 
\begin{equation}
\ddot{\psi}+3H\dot{\psi}=\frac{dU}{d\psi}+\xi R \psi.
\label{eq:7}
\end{equation}
It can be shown that for any value of $\xi$ the phantom behaves like some 
perfect fluid with the effective energy $\rho_{\psi}$ and the pressure 
$p_{\psi}$ in the form which determines the equation of state factor
\begin{equation}
w_{\psi}=\frac{-\frac{1}{2}\dot{\psi}^{2}-U(\psi)
-\xi[2H(\psi^{2})\dot{\ }+(\psi^{2})\ddot{\ }]
-\xi \psi^{2}(2\dot{H}+3H^{2})}{-\frac{1}{2}\dot{\psi}^{2}+U(\psi)
+3\xi H[H\psi^{2}+(\psi^{2})\dot{\ }]} 
\equiv \frac{p_{\psi}}{\rho_{\psi}}.
\label{eq:8}
\end{equation}
Formula (\ref{eq:8}) differs from its counterpart for the standard scalar 
field \cite{Gunzig:2000} by the presence of a negative sign in front of the 
term $\dot{\psi}^{2}$.

The second derivative $(\psi^{2}\ddot{\ }$ in the expression for the pressure 
in eq.~(\ref{eq:8}) can be eliminated and then we obtain
\begin{equation}
p_{\psi}=\left(-\frac{1}{2}-2\xi\right)\dot{\psi}^{2}+\xi H (\psi^{2})\dot{\ }
+ 2\xi(6\xi-1)\dot{H}\psi^{2}+3\xi(8\xi-1)H^{2}\psi^{2}-U(\psi)
+2\xi \psi \frac{dU}{d\psi}.
\label{eq:9}
\end{equation}
Of course such perfect fluid which mimics the phantom field satisfies the 
conservation equation 
\begin{equation}
\dot{\rho}_{\psi}+3H(\rho_{\psi}+p_{\psi})=0.
\label{eq:10}
\end{equation}
We can see that complexity of dynamical equation should manifest by complexity 
of $w_{\psi}$.

Let us consider the FRW quintessential dynamics with some effective energy 
density $\rho_{\psi}$ given in eq.~(\ref{eq:8}). By the quintessence we 
usually understand models with dark energy consisting of a dynamical cosmic 
scalar field. This dynamics can be reduced to the form like of a particle in 
a one-dimensional potential \cite{Szydlowski:2004} and the Hamiltonian of 
the system is 
\begin{equation}
\mathcal{H}(\dot{a},a)=\frac{\dot{a}^{2}}{2}+V(a) \equiv 0, \qquad 
V(a)=-\rho_{\psi}a^{4}.
\label{eq:11}
\end{equation}
The trajectories of the system lie on the zero energy level for flat and 
vacuum models. Note that if we additionally postulate the presence of 
radiation matter for which $\rho_{r} \propto a^{-4}$ then it is equivalent to 
consider the Hamiltonian on the level $\mathcal{H}=E=\text{const}$. Of course 
the division on kinetic and potential parts has only a conventional character 
and we can always translate the term containing $\dot{\psi}^{2}$ into 
a kinetic term.

Let us consider now both case of conformally and minimally coupled phantom 
fields.

\subsection{Conformally coupled phantom fields}

For conformally coupled phantom fields we put $\xi=1/6$ and rescale the field 
$\psi \rightarrow \phi = \psi a$. Then the energy function takes the following 
form for simple mechanical system with a natural Lagrangian function 
$\mathcal{L}=1/2g_{\alpha\beta}\dot{q}^{\alpha}\dot{q}^{\beta}-V(q)$
\begin{equation}
\mathcal{E} = \frac{1}{2}\left(\dot{a}^{2}+\dot{\phi}^{2}\right) 
- \frac{\Lambda}{2}a^{4} - \frac{\lambda}{2}\phi^{4}-m^{2}\phi^{2}a^{2}.
\label{eq:12}
\end{equation}
In contrast to the FRW model with conformally coupled scalar field the kinetic 
energy form is positive definite like for classical mechanical systems. The 
general Hamiltonian which represents the special case of two coupled 
non-harmonic oscillators system is 
\begin{equation}
\mathcal{H}=\frac{1}{2}g^{\alpha\beta}p_{\alpha}p_{\beta}+V(q) 
= \frac{1}{2}(p_{x}^{2}+p_{y}^{2})+Ax^{2}+By^{2}+Cx^{4}+Dy^{4}+Ex^{2}y^{2},
\label{eq:13}
\end{equation}
where $A$,$B$,$C$,$D$, and $E$ are constants.

\subsection{Minimally coupled phantom fields}

For minimally coupled phantom fields ($\xi=0$) the function of energy takes 
the form
\begin{equation}
\mathcal{E}=\frac{\dot{a}^{2}}{2}+\frac{1}{2}(\dot{\phi}a-\phi\dot{a})^{2}
-\frac{\Lambda}{4}a^{4}-\frac{\lambda}{4}\phi^{4}-\frac{1}{2}m^{2}\phi^{2}a^{2}
\label{eq:14}
\end{equation}
where $\rho_{\text{eff}}=-1/2\dot{\psi}^{2}+U(\psi)$, 
$V=-\rho_{\text{eff}} a^{4}$, $\mathcal{H}=1/2\dot{a}^{2}+V(a,\psi,\dot{\psi})$, 
$\phi=a\psi$, $U(\psi)=1/2m^{2}\psi^{2}+1/4\lambda\psi^{4}$ is assumed. 
This time we parameterize the dynamics by taking variable $\psi$ in the 
original cosmological time and the Lagrangian function takes the following form
\begin{equation}
\mathcal{L}=\frac{a'^{2}}{2}+\frac{a^{2}\phi'^{2}}{2}
-\frac{1}{2}m^{2}\phi^{2}a^{2}
-\frac{1}{4}\lambda\phi^{4}a^{2}-\frac{1}{4}\Lambda a^{2},
\label{eq:15}
\end{equation}
where the prime denotes the differentiation with respect to the cosmological 
time parameter $t$, and $V=-\rho_{\text{eff}} a^{2}$, 
$\rho_{\text{eff}}=-1/2\phi'^{2}+U(\phi)$.

\section{Non-integrability as an invariant feature of phantom cosmology with 
scalar field}

For a given Hamiltonian system, it is difficult to show that the system under 
consideration is non-integrable. In general, there are two formulations of 
necessary conditions for the integrability presented by Ziglin 
\cite{Ziglin:1982,Ziglin:1983} and Morales-Ruiz and Ramis 
\cite{Morales:1999,Morales:2001a}. Both approaches base on a deep connection 
between properties of solutions in an enlarged complex time plane and the 
existence of first integrals. This idea originates from works of Kovalevskaya 
and Lyapunov.

Let us study the general case of the Hamiltonian system describing the 
conformally coupled phantom field in the FRW model of the universe. We have
\begin{align}
\label{eq:17}
\mathcal{H}&=\frac{1}{2}(p_{1}^{2}+p_{2}^{2})+V(q_{1},q_{2}), \\
V(q_{1},q_{2})&=\frac{1}{2}\left[\frac{\bar{\Lambda}}{2}q_{1}^{4}
+\frac{\bar{\lambda}}{2}q_{2}^{4}-m^{2}q_{1}^{2}q_{2}^{2}\right], \nonumber
\end{align}
where $q_{1}=a$, $q_{2}=\phi$, $p_{1}=\dot{a}$, $p_{2}=\dot{\phi}$, 
$\bar{\Lambda}=-\Lambda$ and $\bar{\lambda}=-\lambda$. This Hamiltonian has 
the natural form in which the potential is a homogeneous function of degree 
four with respect to both variables $a$, $\phi$. 

From the point of view of complex dynamical behavior it is useful to 
distinguish from Hamiltonian (\ref{eq:17}) with $m^2 = -\mu^2 < 0$. This case 
is interesting because of spontaneous symmetry breaking \cite{Weinberg:1989}. 
The Poincar{\'e} section in this case can be obtained as well as the Lyapunov 
exponents. In this model the chaotic behavior is present. 

In other cases we can define by analogy to Castagnino et al. 
\cite{Castagnino:2000} the monotonic function along trajectories. From this 
fact we obtain that trajectories escape to infinity and the system has no 
recurrence property which guarantee the topological transitivity (the standard 
chaos indicators cannot be obtained). 

Motter and Letelier argued that the cosmological systems with scalar fields 
are non-chaotic but complex in the sense of non-integrability 
\cite{Motter:2002}. Moreover the non-integrability is an invariant property 
of system under the coordinate change. 

In the second distinguished case the complexity has the same character, and 
we apply the some tools to confirm the Liouville non-integrability of this 
system. The Liouville integrability of the Hamiltonian system means that 
there is as many functionally independent functions which are in involution 
(Poisson brackets vanish) as is the dimension of the system.  

Now we consider the problem of non-integrability in both cases. The 
non-integrability of the non-flat first case with the spontaneous symmetry 
breaking was investigated by Ziglin ($\Lambda=\lambda=0$ --- the Yang-Mills 
potential) \cite{Ziglin:2000}. In turn, we apply the Ziglin and Morales-Ruiz 
and Ramis methods to flat phantom models with conformally coupled scalar 
fields with arbitrary parameters. 

The integrability of Hamiltonian systems with a natural Lagrangian was 
analyzed in details by Yoshida 
\cite{Yoshida:1986,Yoshida:1987,Yoshida:1988,Yoshida:1989,Yoshida:1994} 
in the framework of Ziglin's approach. Later Yoshida's results were sharpened 
by Morales-Ruiz and Ramis \cite{Morales:1999}. Note that we applied the 
Morales-Ruiz and Ramis result to system (\ref{eq:17}), but with the indefinite 
kinetic energy form $T=1/2((p_{1}^{2}-p_{2}^{2})$.

The counterpart of Hamiltonian (\ref{eq:17}) for a standard scalar field can 
be obtained after the canonical transformation of variables
\begin{equation*}
q_{1} \rightarrow Q_{1}, \qquad p_{1} \rightarrow P_{1},
\end{equation*}
and
\begin{equation*}
q_{2} \rightarrow iQ_{2}, \qquad p_{2} \rightarrow P_{2}=-ip_{2}.
\end{equation*}
Then of course $dp_{2} \wedge dq_{2} = dP_{2} \wedge dQ_{2}$. However, in this 
case the phase space is complex. Moreover, trajectories have no recurrence 
property which guarantee the topological transitivity, which an essential 
element of the standard understanding of chaos.

The fundamental papers of Ziglin \cite{Ziglin:1982,Ziglin:1983} gave the 
formulation of a very basic theorem about non-integrability of analytic 
Hamiltonian systems. The Ziglin idea connects properties of solutions on a 
complex time plane and the existence of first integrals. This approach takes 
its origins in works of Kovalevskaya and Lyapunov. 

The Yoshida criterion is presented in Appendix. We apply this criterion to 
the analyzed system. Then the equation 
\[
q = V'(q), \qquad q = (q_{1},q_{2})
\]
has the following solutions 
\[
z_{1} = (\pm \bar{\lambda}^{-1/2},0), \quad
z_{2} = (0, \pm \bar{\lambda}^{-1/2}), \quad
z_{3} = \left( \pm \sqrt{\frac{\bar{\lambda} + \mu}{\bar{\Lambda} 
\bar{\lambda} - \mu^{2}}},
\pm \sqrt{\frac{\bar{\lambda} + \mu}{\bar{\Lambda} \bar{\lambda} 
- \mu^{2}}} \right).
\]
The integrability indices for this points are 
\[
\lambda_{i} = - \tr V''(z_{i}) - 3, \qquad i = 1,2,3 
\]
and
\begin{equation}
\label{eq:li}
\lambda_{1} = \frac{\mu}{\Lambda}, \quad 
\lambda_{2} = \frac{\mu}{\lambda}, \quad 
\lambda_{3} = \frac{\lambda_{1} \lambda_{2} - 2(\lambda_{1} + \lambda_{2}) 
+ 3}{1 - \lambda_{1} \lambda_{2}}, \quad
\mu = m^{2}.
\end{equation}
Thus, from the Yoshida criterion follows that if there exists 
$l \in \{ 1,2,3 \}$ such that $\lambda_{l} \in N_{4}$ then system (\ref{eq:17}) 
has no additional meromorphic first integral that is functionally independent 
of $H$. Moreover, our previous application of the Morales-Ruiz and Ramis 
result to the considered system gives that if we introduce quantities 
(\ref{eq:li}) and three discrete sets
\begin{align}
I_{1}=\left\{p(2p-1)\ |\ p \in \mathbb{Z}\right\}, \nonumber \\
I_{2}=\left\{1/8\left[-1+16(1/3+p)^{2}\right]\ |\ p \in \mathbb{Z}\right\}, \nonumber \\
I_{3}=\left\{1/2\left[3/4+4p(p-1)\right]\ |\ p \in \mathbb{Z}\right\},
\end{align}
then if $\lambda_{1},\lambda_{2},\lambda_{3} \notin I=I_{1} \cup I_{2} \cup I_{3}$ the system is 
non-integrable. Therefore, only for certain values of model parameters the phantom cosmology is 
integrable. We can conclude that the Liouville non-integrability is the generic property of the 
system.

If we consider a non-flat model then the effects of curvature are negligible 
near the singularity and the considered case describes a generic situation. 
In this way the phantoms give rise to the complex dynamics in the sense of 
non-existence of a sufficient number of independent first integrals. As a 
consequence, we can express some scepticism about prediction for the equation 
of state factor $w(z)$ in the presence of the phantom component of dark energy. 

Our conclusion is that in a generic case the phantom scalar field can produce 
the complex behavior. The complexity of dynamics is formulated in terms of 
non-integrability (i.e., non-existence of an additional first integral) 
because a standard understanding of chaos has no significant physical meaning 
in the context of a gauge freedom in the choice of a lapse function (time 
parameterization).

Beck \cite{Beck:2004} proposed an interesting idea that stochastically 
quantized scalar fields can offer some solution to the cosmological 
coincidence problem of $\Lambda$. In this approach the chaotic fields have a 
classical equation of state close to $p=-\rho$, i.e., that the chaotic fields 
naturally generate a small cosmological constant. It is possible that phantoms 
are just a phenomenological description of this situation on a purely 
classical level.

Let us also note that regular behavior of dynamics in phantom cosmology can 
appear to be different from the considered types of potentials \cite{Sami:2004} 
but $V(\phi)\propto \phi^{2}$ is the simplest one in which this phenomenon 
occurs. Moreover it can only appear if we treat phantom energy in terms of 
a single scalar field.

\section{Phantom cosmology in terms of barotropic equation of state violating 
the weak energy condition}

It is well-known that for given evolution of the model it is possible to 
construct a potential for a minimally coupled scalar field which would 
reproduce this cosmological evolution \cite{Starobinsky:1998}. Sometimes it 
is possible to find the explicit form of scalar field potential can reproduce 
the evolution arising in some perfect fluid cosmological model 
\cite{Gorini:2004}.

The very different picture is found, if we consider phantom energy as a some 
kind of perfect fluid with super-negative pressure then, in the contrast to 
previous case, the dynamics is regular at very beginning.

Let us consider the dynamics of the FRW models with phantoms where the specific 
form of the equation of state for phantom fluid is assumed. We model the fluid 
which violates the weak energy condition using the equation of state 
$p = w\rho$ and $w = \text{const} < -1$. Such a model of fluid can be treated 
as the simplest phenomenological model of phantom matter. 

The dynamics of this model can be represented by a two-dimensional dynamical 
system (therefore non-chaotic at very beginning) on the phase plane 
$(x,\dot{x}) \equiv (x,y)$ or by motion of a classical particle in the 
one-dimensional potential $V(x)\colon x=a/a_{0}$ \cite{Szydlowski:2004}, i.e.,
\begin{align}
\dot{x}&=y,\nonumber \\
\dot{y}&=-\frac{\partial V}{\partial x}.
\label{fph:1}
\end{align}
System (\ref{fph:1}) has the first integral of energy in the form
\begin{equation}
\frac{\dot{x}^{2}}{2}+V(x)=0,
\label{fph:2}
\end{equation}
where
\begin{equation}
V(x)=-\frac{1}{2}\left(\Omega_{\text{m},0}x^{-1}
+\Omega_{{\text{ph}},0}x^{-3(1+w)+2}+\Omega_{k,0}\right).
\label{fph:3}
\end{equation}
For the mixture of noninteracting matter and phantoms here $\Omega_{i,0}$ are 
the density parameters at the present epoch. In the general case the potential 
of the particle-universe takes the form
\begin{equation}
V(x)=-\frac{1}{2}\sum_{i}\Omega_{i,0}x^{-3(1+w_{i})+2},
\label{fph:4}
\end{equation} 
where $w_{i}=-1$ for the cosmological constant, $w_{i}=-1/3$ for string fluid 
(also curvature fluid), $w_{i}=-2/3$ for topological defects.

\begin{figure}[!ht]
\begin{center}
$\begin{array}{c@{\hspace{0.2in}}c}
\multicolumn{1}{l}{\mbox{\bf (a)}} & 
\multicolumn{1}{l}{\mbox{\bf (b)}} \\ [-0.5cm]
\includegraphics[scale=0.32, angle=270]{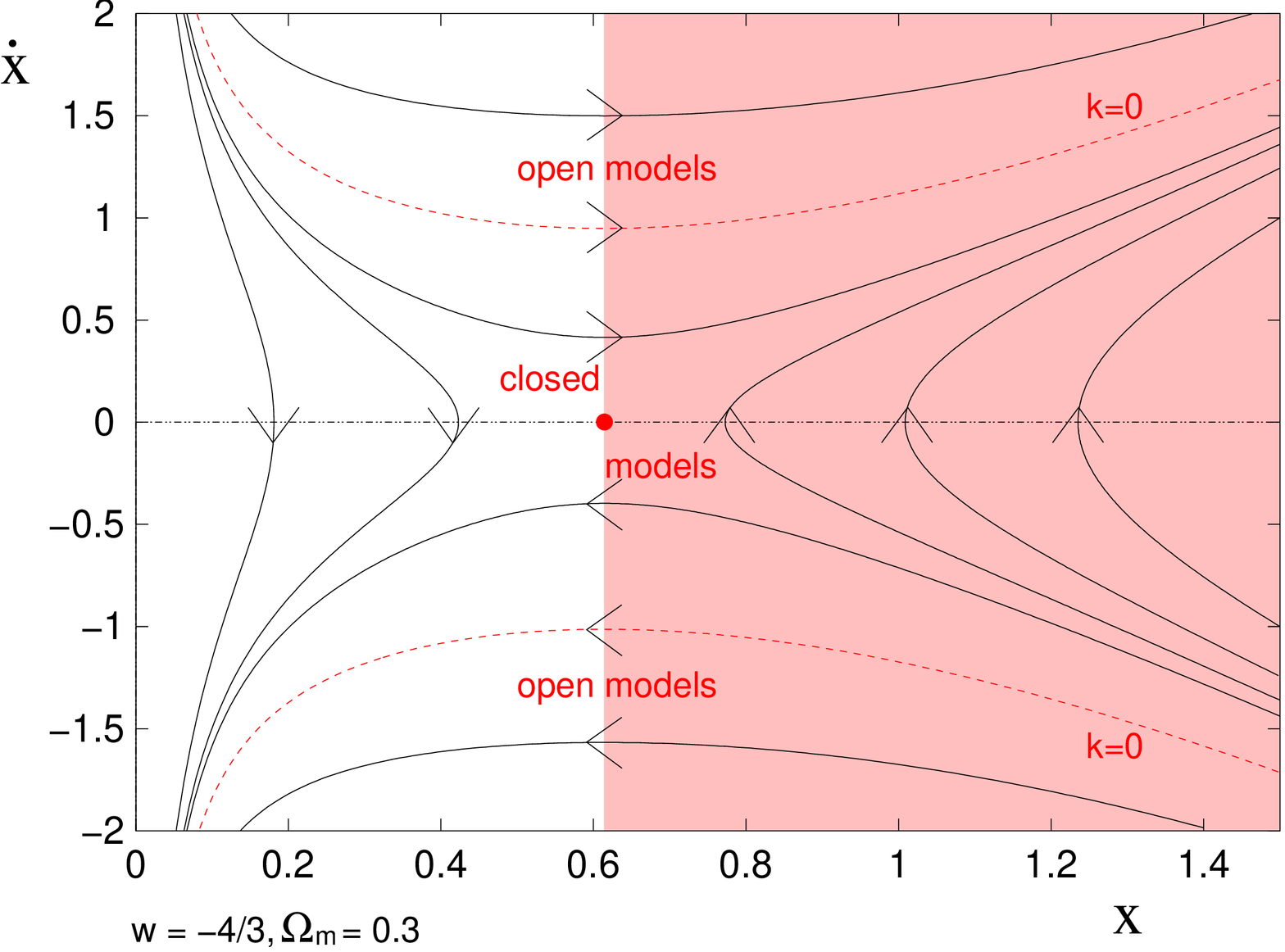} & 
\includegraphics[scale=0.32, angle=270]{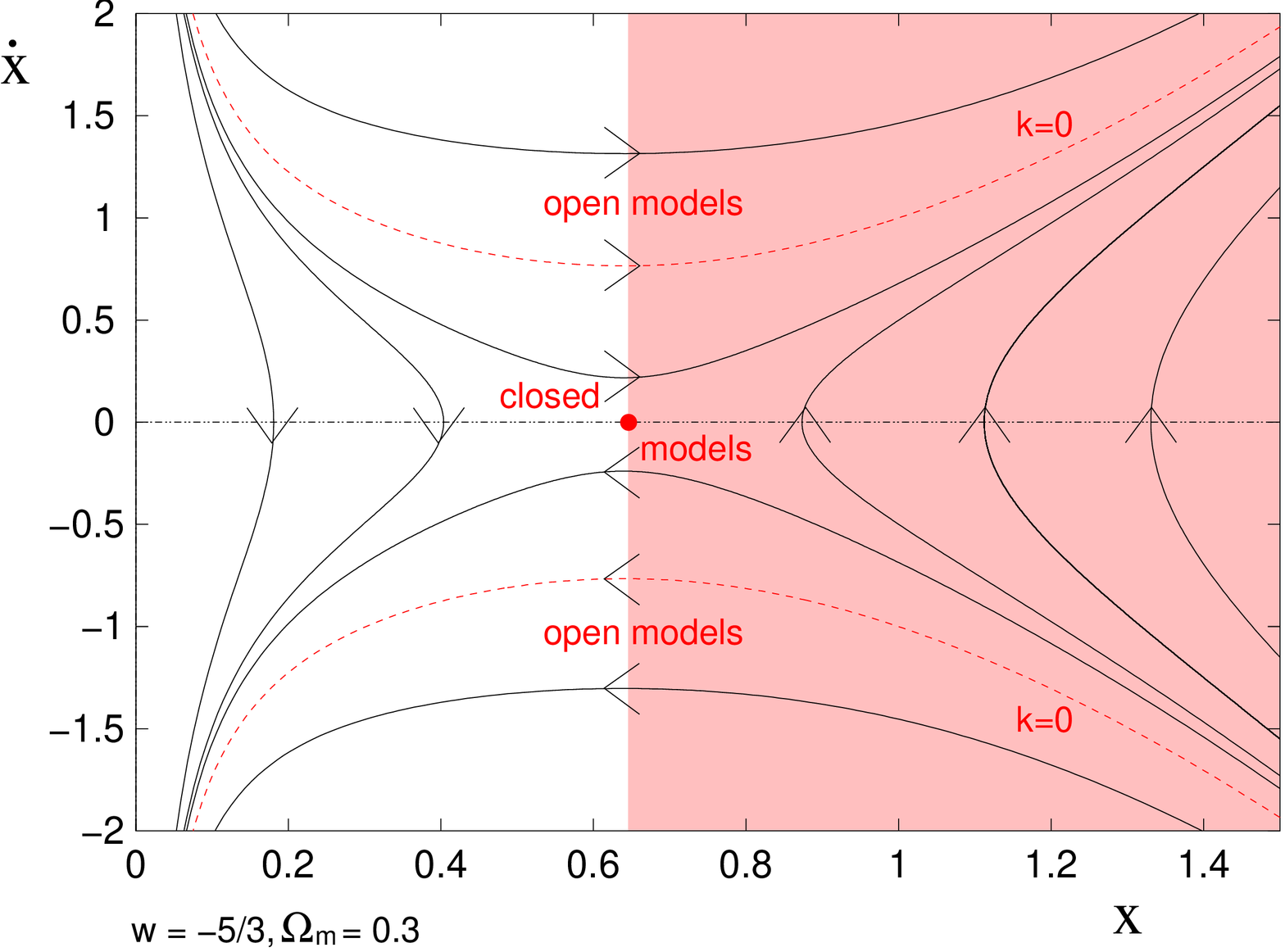} \\ [0.4cm]
\end{array}$
\end{center}
\caption{The phase portraits for the FRW models with phantom matter described 
by the equation of state $p=w\rho$ for (a) $w=-4/3$ and (b) $w=-5/3$. The 
dashed lines are the flat model trajectories. The shaded region is the region 
of accelerated expansion of the universe. Note the topological equivalence 
of both phase portraits.
}
\label{fig1}
\end{figure}

The phase portraits for the model described by system~(\ref{fph:1}) for the 
potential function~(\ref{fph:3}) are shown on Fig.~\ref{fig1}. The trajectory 
of the flat model separates the regions of closed ($\Omega_{k,0}<0$) and open 
($\Omega_{k,0}>0$) models. Moreover, both phase portraits are topologically 
equivalent. The presence of additional terms like strings, topological defects 
(see \cite{Dabrowski:2003}) do not change the structure of the phase plane. 
There is the single critical point located on $x$-axis as an intersection with 
the boundary of the strong energy condition $\rho + 3p \geq 0$. 

Note that the obtained phase portraits are equivalent to phase portraits of 
the FRW model with the cosmological constant.

From the first integral (\ref{fph:2}) for the mixture of the cosmological 
constant and phantom type matter in the flat FRW models we obtain the relation
\begin{equation}
\left(\dot{\ln{x}}\right)^{2}=\left(\frac{\dot{x}}{x}\right)^{2} 
= \Omega_{\text{ph},0}x^{-3(1+w)} + \Omega_{\Lambda,0},
\label{fph:5}
\end{equation}
which preserves its form structure under the change both a position variable 
and a sign of the quintessential parameter $(w+1)$
\begin{equation}
x \rightarrow \frac{1}{x},\qquad (1+w) \rightarrow -(1+w).
\label{fph:6}
\end{equation}
Therefore, if a phantom epoch exists its dynamics can replicate the 
corresponding evolution for the sub-negative equation of state (for example 
$w=-4/3$ corresponds to $w=-2/3$). From this kind of symmetry we obtain that 
if $x(t)$ is the solution of (\ref{fph:5}) for the sub-negative equation of 
state $p=w\rho$ then $x^{-1}(t)$ is also its solution for other form of the 
negative equation of state $p=-(w+2)\rho$. Let us note that for $w=-1$ the 
duality relation which is motivated by superstring theory of duality 
symmetries \cite{Meissner:1991,Lidsey:2000} is the exact symmetry of dynamical 
equations.

\section{Conclusions}

This paper addressed the problem of complexity of the flat FRW dynamics with 
phantom modelled in terms of scalar fields. We proposed a criterion of 
non-integrability in the Liouville sense as an adequate measure of complexity 
of the phantom cosmology. This approach is opted because the gauge freedom in 
the choice of a time parameterization or a lapse function unables us to discuss 
chaos in general relativity in the standard way. Non-integrability is an 
invariant feature of a system and can be use as an indicator of its complex 
dynamics. 

We considered the two approaches to model the phantom fields in the FRW model 
and showed how different dynamics of the models are in these approaches. In 
the first approach which could be called ``microscopic'' we used scalar fields 
in modelling the pressure and energy density. In the second approach we used 
the explicit dependence of pressure versus energy density, which can be 
called phenomenological and found integrable dynamics in contrast to the 
non-integrable dynamics of the first approach. 

In this context non-integrability is a generic feature of the phantom cosmology  
and only for a certain discrete set of values of model parameters phantom 
cosmology is integrable in the Liouville sense. From the physical point of view 
one can interpret this property as the complexity of dynamical behavior of 
trajectories in the phase space. Therefore we cannot expect any simple 
analytical relation for solutions (trajectories) of the system or the form of 
relation $p(\dot{\phi},\dot{a},\phi,a)$ and $\rho(\dot{\phi},\dot{a},\phi,a)$ 
along the trajectories. 

Nevertheless, the integrable cases with zero-measure in the space of all 
solutions exists, they can lead to some analytical dependence of $p(\rho)$, 
after the elimination of time. As the example of simple analytical form of 
the equation of state we considered the barotropic equation of state which 
violates the weak energy condition. We obtained that this model is integrable 
and exhibits the regular dynamics.

We conclude that non-integrability which is a generic feature of the FRW model 
with phantom fields favors rather non-analytical forms of the equation of state. 
However, assuming the barotropic form of equation of state for the phantom 
model we obtain the integrable dynamics at very beginning. We expect that 
phenomenological equation of state can be realized by a microscopic scalar 
field with some potential. Therefore, we assume at very beginning the existence 
of some relation (a first integral) between scalar fields, its derivative and 
evolutional parameter of the universe. From the mathematical point of view 
this requirement means the existence of some invariant in the phase space. 
If we prove non-integrability, than there no such relation in the considered 
class of potentials. If there is even a discrete set of parameters for which 
the system is integrable we hope for finding this relation.

\appendix*

\section{The outline of non-integrability criterion}

Here we present only the facts needed for a formulation of a criterion in 
possibly simple settings. We consider a complex symplectic manifold 
$\mathbb{C}^{2n}$ with the canonical symplectic structure $\Omega$. 
A Hamiltonian vector field $v_H$ is determined by a complex Hamiltonian 
function $H \colon \mathbb{C}^{2n} \longrightarrow \mathbb{C}$ 
by the relation $ \Omega(v_H, \cdot) = dH$.
We assume that Hamilton's analytic equations 
\begin{equation}
\label{eq:ch}
\frac{dz}{dt} = v_H(z), \qquad z=(z_1,\dots, z_{2n})\in\mathbb{C}^{2n}, 
\quad t \in \mathbb{C},
\end{equation}
have the non-equilibrium solution  $z=\varphi(t)$. To simplify the exposition 
we assume that this solution lies on a two-dimensional invariant plane 
\[
\Pi =\{(z_1,\dots, z_{2n}) \in \mathbb{C}^{2n} \, | \, z_{i}=0, 
\quad i=1,\ldots, 2(n-1)\}.
\]
The phase curve $\Gamma = \{ \varphi(t)\in\mathbb{C}^{2n}\,|\, 
t \in \mathbb{C} \}$ is a Riemannian surface with a local coordinate~$t$. 
Together with equations (\ref{eq:ch}) we consider also variational equations 
along solution $\varphi(t)$
\begin{equation}
\label{eq:cv}
\frac{d \xi}{dt} = A(t) \xi, \qquad 
A(t) = \frac{\partial v_H}{\partial z}(\varphi(t)). 
\end{equation}
This system separates into the normal and the tangential subsystems. 
In our settings this separation takes a very simple form---the matrix $A(t)$ 
has a block diagonal structure.  We consider the normal variational 
equations (NVE) 
\begin{equation}
\label{eq:cn}
\frac{d \eta}{dt} = B(t) \eta,  \qquad \eta\in \mathbb{C}^{2(n-1)},
\end{equation}
where $B(t)$ is $2(n-1)\times 2(n-1)$ upper diagonal block of matrix $A(t)$.
We choose a point $t_0\in\mathbb{C}$ and a matrix of fundamental solutions 
of the NVE $X(t)$, defined in a neighborhood of $t_0$. With a close path 
$\alpha$ on complex time plane starting and ending at point $t_0$ we can 
associate a matrix $S\in \text{GL}(2(n-1),\mathbb{C})$ in the following 
way. We integrate NVE (\ref{eq:cn}) along the path $\alpha$, i.e., 
we make an analytic continuation of $X(t)$ along this path. As a result 
from the fundamental solution $X(t)$ we obtain another fundamental 
solution $Y(t)$. From the general theory of linear systems it follows 
that $Y(t)=SX(t)$ for some $S\in \text{GL}(2(n-1),\mathbb{C})$. Because 
the system is Hamiltonian, $S$ is a symplectic matrix, i.e, 
$S\in \text{Sp}(2(n-1),\mathbb{C})$. 
In this way, considering all possible paths, we obtain a matrix 
representation of the first homotopy group $\pi_1(\Gamma)$ of $\Gamma$. 
It forms a finitely generated subgroup of $\text{Sp}(2(n-1),\mathbb{C})$ 
and it is called a monodromy group. We denote it $M$. 

Let us take an element of the monodromy group $g\in M$. Its spectrum has the 
form 
\[
\text{spectr} (g) = ( \lambda_1, \lambda_1^{-1}, \ldots, 
\lambda_{n-1}, \lambda_{n-1}^{-1}), \quad \lambda_i \in\mathbb{C}.
\]
The element $g$ is resonant if 
\[
\prod_{l=1}^{n-1} \lambda_l^{k_l} =1\ \text{for some} \ 
(k_1, \ldots, k_{n-1})\in \mathbb{Z}^{n-1}\backslash\{0\}.
\]
\begin{theorem}[Ziglin \cite{Ziglin:1982}] Let us assume that there exists a 
non-resonant element $g\in M$. If the Hamiltonian system possesses 
in a connected neighborhood of $\Gamma$ $n-1$ meromorphic first 
integrals which are functionally independent with $H$ then 
for an element $g'\in M$: if  $g e = \lambda e$ for 
$\lambda\in\mathbb{C}$ and $e\in\mathbb{C}^{2(n-1)}$, 
then $g (g'e)= \lambda' (g'e)$ for some $\lambda'\in\mathbb{C}$.  
\end{theorem}

In the case of a system with two degrees of freedom this theorem can be 
formulated in a more operational way. 
\begin{theorem}
Let us assume that there exists a non-resonant element $g\in M$. 
If there exists other element $g'\in M$ such that
\begin{enumerate}
 \item $\tr g'\neq 0$ and $gg'\neq g'g$, or
 \item $\tr g'= 0$ and $gg'g\neq g'$,
\end{enumerate}
then there is no additional meromorphic first integral functionally independent 
of $H$ in a connected neighborhood of $\Gamma$.
\end{theorem}

The main difficulty with the application of the Ziglin theorem is the 
determination of the monodromy group of the NVE. Only in very special 
cases we can do this analytically. Yoshida 
\cite{Yoshida:1986,Yoshida:1987,Yoshida:1988,Yoshida:1989} developed the Ziglin 
approach for these cases when the Hamiltonian of a system has the 
natural form and the potential is a homogeneous function. In this case 
a particular solution can be found in the form of `straight line 
solution' and the NVEs for it can be transformed to a product of 
certain copies of hyper-geometric equations for which the monodromy 
group is known. This allows to formulate adequate theorems in a form 
of an algorithm. Below we describe it for the Hamiltonian system with 
two degrees of freedom.  

Consider the Hamiltonian 
\begin{equation}
\label{eq:hY}
H = \frac{1}{2} (p_{1}^{2} + p_{2}^{2}) + V(q_1,q_2), \qquad 
(q_1,q_2,p_1,p_2)\in\mathbb{C}^4, 
\end{equation}
where $V(q_1,q_2)$ is the homogeneous function of degree $k$, i.e.,
\begin{equation}
\label{eq:v}
V(Cq_1,Cq_2) = C^{k} V(q_1,q_2).
\end{equation}
In a generic case this system has a straight line solutions of the form 
\begin{equation}
\label{eq:sls}
q_1 = C_{1} \phi(t), \qquad 
q_2 = C_{2} \phi(t)
\end{equation}
where $\phi(t)$ is a solution of a nonlinear equation 
\[
\ddot{\phi} = - \phi^{k-1}
\]
and $(C_{1}, C_{2})\neq(0,0)$ are solutions of the following system 
\begin{equation}
\label{eq:c}
C_{1} = \partial_{1} V(C_{1},C_{2}), \quad
C_{2} = \partial_{2} V(C_{1},C_{2}). 
\end{equation}

The variational equations take the form 
\[
\left[ \begin{array}{c} 
\ddot{\xi} \\ \ddot{\eta} 
\end{array} \right] 
= - \left[ \begin{array}{cc} 
V_{11} & V_{12} \\ V_{21} & V_{22} 
\end{array} \right] 
\left[ \begin{array}{c} 
\xi \\ \eta 
\end{array} \right] 
(\phi(t))^{k-2},
\] 
where $V_{ij}= \partial_i\partial_j V(C_1,C_2)$ for $i,j=1,2$.
Since the Hessian of $V$ is symmetric it is diagonalizable 
by an orthogonal transformation and the system separates to 
\begin{align}
\ddot{\xi} &= - \lambda_{1} \Phi^{k-2}(t) \xi, \\
\ddot{\eta} &= - \lambda_{2} \Phi^{k-2}(t) \eta, \label{eq:nve2}
\end{align}
where $\lambda_{1}$, $\lambda_{2}$ are real eigenvalues of 
the Hessian. Let us note that it is not true for indefinite 
systems where the Hessian is not a symmetric matrix. 

It can be shown that the Hessian of $V$ at $C=(C_{1},C_{2})$ has the eigenvalue 
$\lambda_{1} = k-1$.  Thus, its second eigenvalue is equal 
$\lambda_{2} = \tr V(C_{1},C_{2}) - (k-1)$, and it is called the integrability 
index. Equation (\ref{eq:nve2}) can be transformed to the hyper-geometric 
equation. The monodromy matrices of this equation are parameterized by 
$\lambda$ and conditions of the Ziglin theorem put a restriction on values of 
$\lambda$---simply, we can identify those values of $\lambda$ for which the 
system is not integrable (more precisely: it does not possess an additional 
meromorphic first integral). To state it accurately let us define 
\begin{equation}
\label{eq:ik}
I_{k}(p) = \left[ \frac{k}{2} p(p+1) - p , 
\frac{k}{2} p(p+1) + p + 1 \right] , \quad 
p \in \mathbb{N},
\end{equation}
and
\begin{equation}
\label{eq:nk}
N_{k} = \mathbb{R} \setminus \bigcup_{p \in \mathbb{N}} I_{k}(p).
\end{equation}
Then it follows that Hamiltonian system (\ref{eq:hY}) with homogeneous
potential (\ref{eq:v}) of degree $k$ is not integrable if the integrability
index $\lambda$ corresponding to a certain straight line solution 
(\ref{eq:sls}) belongs to $N_k$. Let us note that equations (\ref{eq:c}) 
usually have several solutions and thus it is necessary to check the Yoshida 
criterion for each of them.

\begin{acknowledgments}
The authors are very grateful to the anonymous referee for thoughtful remarks 
and Orest Hrycyna for discussion and comments.
The paper was supported by KBN grant 1 P03D 003 26. 
\end{acknowledgments}

\end{document}